\pdfoutput=1
\documentclass[runningheads,a4paper,envcountsame]{article}

\usepackage{amsmath,amsgen,latexsym}
\usepackage{amstext,amssymb,amsfonts,latexsym}
\usepackage{theorem}
\usepackage{pifont}
\usepackage{graphicx}
\usepackage{listings}
\usepackage{relsize}

 \newcommand{\bs}{\bigskip}
 \newcommand{\ms}{\medskip}
 \newcommand{\n}{\noindent}
 \newcommand{\s}{\smallskip}
 \newcommand{\hs}[1]{\hspace*{ #1 mm}}
 \newcommand{\vs}[1]{\vspace*{ #1 mm}}

 \newcommand{\setempty}{\varnothing}
 \newcommand{\real}{\mathbb{R}}
 \newcommand{\nat}{\mathbb{N}}
 
 \newcommand{\integer}{\mathbb{Z}}

 \newcommand{\CC}{{\cal C}}
 \newcommand{\FF}{{\cal F}}

 \newcommand{\LL}{{\cal L}}

 \newcommand{\SSS}{{\cal S}}
 
 \newcommand{\PP}{{\cal P}}

 \newcommand{\dl}{\mathrm{L}}
 \newcommand{\nl}{\mathrm{NL}}
 \newcommand{\p}{\mathrm{P}}
 \newcommand{\np}{\mathrm{NP}}

 \newcommand{\fl}{\mathrm{FL}}

 \newcommand{\reg}{\mathrm{REG}}
 \newcommand{\cfl}{\mathrm{CFL}}

\theoremstyle{plain}
\theoremheaderfont{\bfseries}
 \newtheorem{theorem}{Theorem}[section]
 \newtheorem{lemma}[theorem]{Lemma}
 \newtheorem{proposition}[theorem]{{\bf Proposition}}
 \newtheorem{corollary}[theorem]{Corollary}

 {\theorembodyfont{\rmfamily}
  }
 {\theorembodyfont{\rmfamily} }
 {\theorembodyfont{\rmfamily} }

 \newtheorem{claim}{Claim}

 \newenvironment{proofsketch}{\par \noindent
            {\bf Proof Sketch. \hs{2}}}{\hfill$\Box$ \vspace*{3mm}}

 \newcommand{\floors}[1]{\lfloor #1 \rfloor}
 \newcommand{\pair}[1]{\langle #1 \rangle}

\newcommand{\ignore}[1]{}

   \newcommand{\logcfl}{\mathrm{LOGCFL}}
 
 \newcommand{\shift}{\mathrm{shift}\mbox{-}}

 \newcommand{\Lmreduces}{\leq^{\mathrm{L}}_{m}}
 \newcommand{\Lttreduces}{\leq^{\mathrm{L}}_{tt}}

 \newcommand{\auxpdatisp}{\mathrm{AuxPDATI,\!SP}}

\begin{document}

\pagestyle{plain}
\pagenumbering{arabic}
\setcounter{page}{1}
\setcounter{footnote}{0}

\begin{center}
{\Large {\bf When Input Integers are Given in the Unary \s\\ Numeral Representation}}\footnote{This is a preliminary report of the current work, which has appeared in the Proceedings of the 24th Italian Conference on Theoretical Computer Science (ICTCS 2023), Palermo, Italy, September 13--15, 2023, CEUR Workshop Proceedings (CEUR-WS.org).}
\bs\s\\

{\sc Tomoyuki Yamakami}\footnote{Present Affiliation: Faculty of Engineering, University of Fukui, 3-9-1 Bunkyo, Fukui 910-8507, Japan} \bs\\
\end{center}
\ms


\begin{abstract}
Many $\np$-complete problems take integers as part of their input instances. These input integers are generally binarized, that is, provided in the form of the ``binary'' numeral representation, and the lengths of such binary forms are used as a basis unit to measure the computational complexity of the problems.
In sharp contrast, the ``unarization'' (or the ``unary'' numeral representation) of numbers has been known to bring a remarkably different effect onto the computational complexity of the problems. When no computational-complexity difference is observed   between binarization and unarization of instances, on the contrary, the problems are said to be strong NP-complete.
This work attempts to spotlight an issue of how the unarization of instances affects the computational complexity of various combinatorial problems.
We present numerous NP-complete (or even NP-hard) problems, which turn out to be easily solvable when input integers are represented in unary.
We then discuss the computational complexities of such problems when taking unary-form integer inputs. We hope that a list of such problems signifies the structural differences between strong NP-completeness and non-strong NP-completeness.

\s
\n{\bf Keywords.} 
  unary numeral representation, 
  unarization of integers, 
  logarithmic-space computation, 
  pushdown automata, 
  log-space reduction
\end{abstract}


\section{Background and Quick Overview}\label{sec:introduction}

\subsection{Unary Representations of Integer Inputs}\label{sec:L-vs-NL}

The \emph{theory of NP-completeness} has made great success in providing a plausible evidence to the hardness of target computational problems when one tries to solve them in feasible time. The proof of $\np$-completeness of those problems therefore makes us turn away from solving them in a practical  way but it rather guides us to look into the development of  approximation or randomized algorithms.

In computational complexity theory, we often attempt to determine the minimum amount of computational resources necessary to solve target combinatorial problems. Such computational resources are measured in terms of the sizes of input instances given to the problems.
Many $\np$-complete problems, such as the \emph{knapsack problem}, the \emph{subset sum problem}, and the \emph{integer linear programming problem}, concern the values of integers and require various integer manipulations.
When instances contain integers, these integers are usually expressed in the form of ``binary'' representation.
Thus, the computational complexities, such as running time or memory space, are measured with respect to the total number of bits used for this representation.

This fact naturally brings us a question of what consequences are drawn when input integers are all provided in ``unary'' using the unary numeral system instead.
How does the unary representation affects the computational complexity of the   problems?
In comparison to the binary numeral system, the unary numeral system is so distinctive,  that we need to heed a special attention in our study on the computational complexity of algorithms.

When input integers given to combinatorial problems are expressed in unary, a simple transformation of the unary expression of input integers to their binary representations makes the original input lengths look exponentially larger than their binary lengths. Thus, any algorithm working with the unarized integer inputs seem to consume exponentially less time than the same algorithm with binarized integer inputs. This turns out to be a quite short-sighted analysis.

We often observe that the use of the unary representation significantly alters the computational complexity of combinatorial problems.
However, a number of problems are known to remain $\np$-complete even after we switch  binarized integer inputs to their corresponding unarized ones (see \cite[Section 4.2]{GJ79}). Those problems are said to be \emph{strong NP-complete}\footnote{Originally, the strong $\np$-completeness has been studied in the case where all input integers are \emph{polynomially bounded}. This case is essentially equivalent to the case where all input integers are given in unary. See a discussion in, e.g., \cite[Section 4.2]{GJ79}.}
and this notion
has been used to support a certain aspect of the robustness of $\np$-completeness notion. Non-strong  $\np$-complete problems are, by definition, quite susceptible to the change of numeral representations of their input integers between the binary representation and the unary representation.
It is therefore imperative to study  a structural-complexity aspect of those non-strong $\np$-complete problems
when input integers are provided in the form of the unary numeral representation.
In this work, we wish to look into the features of such non-strong $\np$-complete problems.

Earlier, Cook \cite{Coo85} discussed the computational complexity of a unary-representation analogue of the knapsack problem, called the \emph{unary 0-1 knapsack problem} (UK), which asks whether or not there is a subset of given positive integers, represented in unary, whose sum matches a given target positive integer. This problem UK naturally falls in $\nl$ (nondeterministic logarithmic space) but he conjectured that UK may not be $\nl$-complete.
This conjecture is supported by the fact that even an appropriately designed one-way one-turn nondeterministic counter automaton can recognize $\mathrm{UK}$ (e.g., \cite{CH88}).
Jenner \cite{Jen95} introduced  a variant of UK whose input integers are given in a ``shift-unary'' representation, where a  \emph{shift-unary representation}  $[1^a,1^b]$ represents the number $a\cdot 2^b$. She then demonstrated that this variant is actually $\nl$-complete.
We further expand such a shift-unary representation to a \emph{multiple shift-unary representation} by allowing a finite series of positive integers
and to a \emph{general unary (numeral) representation} by taking all possible  integers, including zero and negative ones.
See Section \ref{sec:numbers} for their precise definitions.

Driven by our great interest in the effect of the unarization of inputs, throughout this work, we wish to study the computational complexity of combinatorial problems whose input integers are in part unarized by the unary representation.

For the succinctness of further descriptions, we hereafter refer to input integers expressed in the unary numeral system as \emph{unary-form integers} (or more succinctly, \emph{unarized integers}) in comparison to \emph{binary-form integers} (or \emph{binarized integers}).

\subsection{New Challenges and Main Contributions}\label{sec:main-contribution}

A simple pre-processing of converting a given unary-form input integer into its binary representation provides obvious complexity upper bounds for target combinatorial problems but it does not seem to be sufficient to determine their precise computational complexity.

Our goal is to explore this line of study in order to clarify the roles of the binary-to-unary transformation in the theory of $\np$-completeness (and beyond it) and cultivate a vast research area incurred by the use of unary-form  integers.
In particular, we plan to focus on several non-strong $\np$-complete (or even NP-hard) problems and study how their computational complexities can change when we replace the binary-form input integers to the unary-form ones.
Among various types of combinatorial problems taking integer inputs, we look into number problems, graph-related problems, and lattice-related problems.
It is desirable for us to determine the precise complexity of each combinatorial problem whose instances are unarized.

\begin{figure}[t]
\centering
\includegraphics*[height=4.8cm]{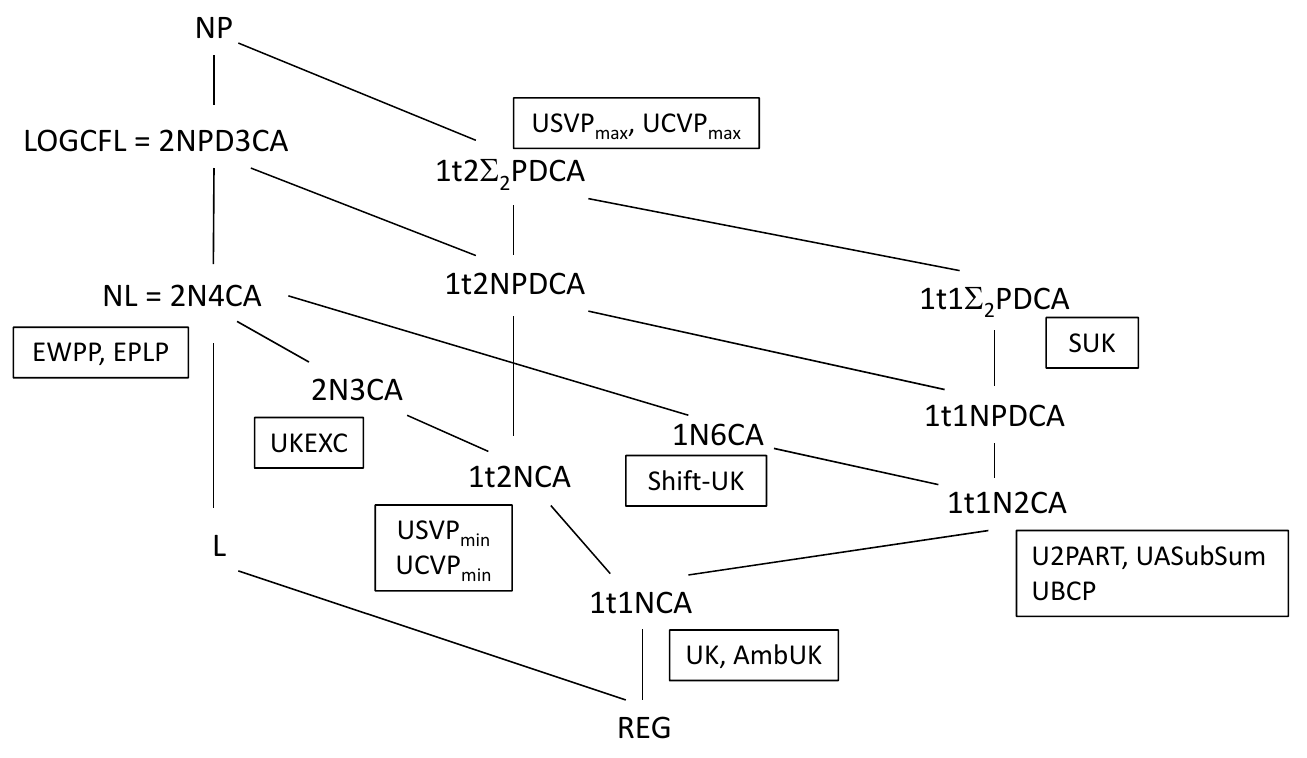}
\caption{Inclusion relationships among complexity classes with solid lines and membership relations of numerous decision problems listed in small boxes to specific complexity classes.}\label{fig:2CVC-graph-figure}
\end{figure}

A brief summary of our results are illustrated in Figure \ref{fig:2CVC-graph-figure}. The detailed explanation of the combinatorial problems and complexity classes listed in this figure will be provided in Sections \ref{sec:preparation}--\ref{sec:lattice-theory}. We remark that the underlying machines used to define those complexity classes are all limited to run in polynomial time.

\section{Basic Notions and Notation}\label{sec:preparation}

\subsection{Numbers, Sets, Languages}\label{sec:numbers}

We assume that all \emph{polynomials} have nonnegative integer coefficients. All \emph{logarithms} are always taken to the base $2$.
The notations $\integer$ and $\nat$ denote respectively the set of all integers and that of all nonnegative integers. We further define $\nat^{+}$ to be  $\nat-\{0\}$.
As a succinct notation, we use $[m,n]_{\integer}$ to denote the \emph{integer interval}  $\{m,m+1,\ldots,n\}$ for two integers $m$ and $n$ with $m\leq n$. In particular, $[1,n]_{\integer}$ is abbreviated as $[n]$ whenever $n\geq1$. 
Given $n\in\nat^{+}$, we define the linear order $<_{[n]}$ on the set $[n]\times[n]$ as: $(i_1,i_2)<_{[n]} (i_2,j_2)$ iff either $i_1<i_2$ or $i_1=i_2 \wedge j_1<j_2$.
Moreover, $\real$ denotes the set of all real numbers.
Given a vector $x=(x_1,x_2,\ldots,x_n)$ in $\real^n$, the \emph{Euclidean norm} $\|x\|_{2}$ of $x$ is given by $(\sum_{i\in[n]} x_i^2)^{1/2}$ and the \emph{max norm} $\|x\|_{\infty}$ is done by $\max\{|x_i|: i\in[n]\}$, where $|\cdot|$ indicates the absolute value. Given a set $A$, $\PP(A)$ denotes the \emph{power set} of $A$.

Conventionally, we freely identify decision problems with their associated languages. We write $1^*$ (as a regular expression) for the set of strings of the form $1^n$ for any $n\in\nat$. For convenience, we define $1^0$ to be the \emph{empty string} $\varepsilon$. Let $1^{+}$ denote the set $1^*-\{\varepsilon\}$. Similarly, we use the notation $0^*$ for $\{0^n\mid n\in\nat\}$ with $0^0=\varepsilon$.

Given a positive integer $a$, the \emph{unary representation} of $a$ is of the form $1^a$ (viewed as a string) compared to its binary representation.
Notice that the length of $1^a$ is exactly $a$ rather than $O(\log(a+1))$,  which is the length of the binary representation of $a$.
A finite series $(a_1,a_2,\ldots,a_n)$ of positive integers is expressed by the \emph{multiple unary representation} of the form $(1^{a_1},1^{a_2},\ldots,1^{a_n})$.
When such an instance $x=(1^{a_1},1^{a_2},\ldots,1^{a_n})$ is given to a machine, we explicitly assume that $x$ has the form of $1^{a_1}\# 1^{a_2}\# \cdots \# 1^{a_n}$ with a designated separator symbol $\#$.
For a series of such instances $x=(x_1,x_2,\ldots,x_m)$ with $x_i=(1^{a_{i1}},1^{a_{i2}},\ldots,1^{a_{in_i}})$ for $i\in[m]$ and $n_i\in\nat^{+}$, we further assume that $x$ takes the form of $1^{a_{11}}\# 1^{a_{12}} \#\cdots\# 1^{a_{1n_1}} \#\#  1^{a_{21}}\# 1^{a_{22}}\# \cdots \#1^{a_{2n_2}} \#\# \cdots\#\# 1^{a_{m1}} \# 1^{a_{m2}}\# \cdots\# 1^{a_{mn_m}}$.
Moreover, for any positive integer of the form $a=p\cdot 2^t$ for nonnegative integers $p$ and $t$, a \emph{shift-unary representation}\footnote{Unlike the unary and binary representations, a positive integer in general has more than one shift-unary representation. In most applications, the choice of such representations does not significantly affect the computational resources of running machines.} of $a$ is a pair $[1^p,1^t]$, which is different from the unary representation $1^a$ of $a$. The length of $[1^p,1^t]$ is $O(p+t)$ but not $a$.
We also use a \emph{multiple shift-unary representation}, which has the form $[[1^{a_1},1^{b_1}], [1^{a_2},1^{b_2}], \ldots, [1^{a_n},1^{b_n}]]$  with the condition that $2^{b_{i+1}}>a_i2^{b_i}$ for all $i\in[n-1]$. This form represents the number $\sum_{i=1}^{n}a_i\cdot 2^{b_i}$.
We intend to call an input integer by the name of its representation. For this purpose, we call the expression $1^a$ and $[1^p,1^t]$ the  \emph{unary-form (positive) integer} (or more succinctly, the \emph{unarized (positive) integer}) and the \emph{shift-unary-form (positive) integer}, respectively.

To deal with ``general'' integers, including zero and negative integers, however, we intend to express such an integer $a$ as a unary string by applying the following special encoding function  $\pair{ \cdot }$. Let $\pair{a}=1$ if $a=0$; $\pair{a}=2a$ if $a>0$; and $\pair{a}=2|a|+1$ if $a<0$. We define the \emph{general unary (numeral) representation} of $a$ as $1^{\pair{a}}$. In a similar way,
a \emph{general shift-unary representation} of $-a$ for $a>0$ is defined as a pair of the form  $[1^{\pair{-p}},1^{\pair{t}}]$, where $p$ and $t$ in $\nat$ must satisfy $a = p\cdot 2^t$.

\subsection{Turing Machines and Log-Space Reductions}

Our interest mostly lies on space-bounded computation. As a basic machine model, we thus use deterministic/nondeterministic Turing machines (or DTMs/NTMs, for short), each of which is equipped with a read-only input tape, a rewritable work tape, and (possibly) a write-once\footnote{A tape is called \emph{write-once} if its tape head never moves to the left and, whenever it writes a non-blank symbol, it must move to the next blank cell.} output tape.

The notation $\dl$ (resp., $\nl$) denotes the collection of all decision problems solvable on DTMs (resp., NTMs) in polynomial time using logarithmic space (or log space, for short). A function $f$ from $\Sigma^*$ to $\Gamma^*$ for two alphabets $\Sigma$ and $\Gamma$ is \emph{computable in log space} if there is a DTM such that, on input $x$, it produces $f(x)$ on a write-once output tape in $|x|^{O(1)}$ time and $O(\log|x|)$ space. The notation $\fl$ refers to the class of all such functions.

Let $L_1$ and $L_2$ denote two arbitrary languages. We say that $L_1$ is \emph{$\dl$-m-reducible to $L_2$} (denoted $L_1\Lmreduces L_2$) if there exists a function $f$ (called a reduction function) in $\fl$ such that, for any $x$, $x\in L_1$ iff $f(x)\in L_2$.
Given a language family $\FF$, the notation $\mathrm{LOG}(\FF)$ denotes the family of all languages that are $\dl$-m-reducible to appropriately chosen languages in $\FF$.
We further say that $L_1$ is \emph{$\dl$-tt-reducible to $L_2$} (denoted $L_1\Lttreduces L_2$) if there are a reduction function $f\in\fl$ and a truth-table $E: \{0,1\}^* \to \{0,1\}$ in $\fl$ such that, for any string $x$, (i) $x\in L_1$ iff $f(x)=(y_1,y_2,\ldots,y_m)$ with $y_i\in \Sigma^*$ for any index $i\in[m]$ and (ii) $E(L_2(y_1),L_2(y_2), \ldots,L_2(y_m)) =1$, where $L_2(y)=1$ if $y\in L_2$ and $L_2(y)=0$ otherwise.

Before solving a given problem on an input $(1^{a_1},1^{a_2},\ldots,1^{a_n})$ of unary-form numbers, it is often useful to sort all entries $(a_1,a_2,\ldots,a_n)$ of this input.
Let us define the function $f_{order}$ making the following behavior: on input of the form $(1^{a_1},1^{a_2},\ldots,1^{a_n})$ with $a_1,a_2,\ldots,a_n\in\nat^{+}$, $f_{order}$ produces a tuple $(1^{a_{i_1}},1^{a_{i_2}},\ldots,1^{a_{i_n}})$ such that (1) $a_{i_1}\geq a_{i_2}\geq \cdots \geq a_{i_n}$ and (2) if $a_{i_j}=a_{i_k}$ with $i_j\neq i_k$, then $i_j<i_k$ holds. Condition (2) is a useful property for one-way finite automata.

There is an occasion where, for a set of shift-unary-form integers $[1^{p_1},1^{t_1}],[1^{p_2},1^{t_2}],\ldots, [1^{p_n},1^{t_n}]$, we wish to compute the sum $s= \sum_{i=1}^{n} p_1\cdot 2^{t_i}$ and output the binary representation of $s$ in the reverse order. The notation $f_{sum}$ denotes the function that computes this value $s$. The following lemma is useful.

\begin{lemma}\label{function-sum}
The functions $f_{order}$ and $f_{sum}$ are both in $\fl$.
\end{lemma}

Those functions will be implicitly used for free when solving combinatorial problems in the subsequent sections.

\subsection{Multi-Counter Pushdown Automata}

A \emph{one-way determinsitic/nondeterminitic pushdown automaton} (or a 1dpda/1npda, for short) is another computational model with a read-once input tape and a standard (pushdown)  stack whose operations are restricted to the topmost cell.
A \emph{counter} is a FILO (first in, last out) memory device that behaves like a stack but its alphabet consists only of a ``single'' symbol, say, $1$ except for the bottom marker $\bot$.
A \emph{one-way nondeterministic $k$-counter automaton} (or a $k$-counter 1ncta) is a one-way nondeterministic finite automaton (1nfa) equipped with $k$ counters.
We write $\mathrm{1N}k\mathrm{CA}$ to denote the family of
all languages recognized by appropriate $k$-counter 1ncta's running in polynomial time.
We further expand $\mathrm{1N}k\mathrm{CA}$ to $\mathrm{1NPD}k\mathrm{CA}$ by supplementing $k$ counters to 1npda's.
These specific machines are called \emph{one-way nondeterministic $k$-counter pushdown automata} (or 1npdcta's).
When tape heads of multi-counter automata and pushdown automata are allowed to move in all directions, we call such polynomial-time machines by 2ncta's and 2npdcta's, respectively. With the use of these 2ncta's and 2npdcta's, we respectively obtain $\mathrm{2N}k\mathrm{CA}$ and $\mathrm{2NPD}k\mathrm{CA}$.

An \emph{alternating machine} must use two groups of inner states: existential states and universal states.
We are concerned with the number of times that such a machine  switches between existential and universal inner states.  When this number is upper-bounded
by $k$ ($k\geq0$) along all computation paths of $M$ on any input $x$, the machine is said to have at most $k+1$ \emph{alternations}.
For any $k\geq1$, the complexity class $1\Sigma_k\mathrm{PDCA}$ (resp., $2\Sigma_k\mathrm{PDCA}$) is composed of all languages recognized by \emph{one-way} (resp., \emph{two-way}) \emph{alternating 1-counter pushdown automata} running in polynomial time with at most $k$ alternations starting with existential inner states. Note that $1\Sigma_1\mathrm{PDCA}$ coincides with $\mathrm{1NPDCA}$.

The notion of \emph{turns} was studied initially by Ginsburg and Spanier \cite{GS66}. Turn-restricted counter automata were called \emph{reversal bounded} in the past literature. A 1ncta is said to \emph{make a turn} along a certain accepting computation path if the stack height (i.e., the size of stack's content) changes from nondecreasing to decreasing exactly once.
A \emph{1-turn 1ncta} is a 1ncta that makes at most one turn during each computation. We add the prefix ``1t'' to express the restriction that every underlying machine makes at most one turn.
For example, we write $\mathrm{1t1NCA}$ when all underlying 1ncta's in the definition of $\mathrm{1NCA}$ are 1-turn 1ncta's. Similarly, we define, e.g., $\mathrm{1t2NPDCA}$ and $\mathrm{1t2}\Sigma_k\mathrm{PDCA}$ by requiring all stacks and counters to make at most one turn. It follows that $\reg\subseteq \mathrm{1t1NCA}\subseteq \mathrm{1NCA} \subseteq \cfl$, where $\reg$ (resp., $\cfl$) denotes the class of all regular (resp., context-free) languages. Notice that $\cfl=\mathrm{1NPD}$.
It also follows that $\dl\subseteq \mathrm{LOG}(\mathrm{1t1NCA})\subseteq \mathrm{LOG(1NCA)} = \nl$. Conventionally, we write $\logcfl$ instead of  $\mathrm{LOG}(\cfl)$.

\begin{lemma}\label{counter-inclusion}
For any $k\geq1$, $\mathrm{2N}k\mathrm{CA} \subseteq \nl$, $\mathrm{2NPD}k\mathrm{CA} \subseteq \logcfl$, and $2\Sigma_2\mathrm{PD}k\mathrm{CA}\subseteq \np$.
\end{lemma}

\begin{proofsketch}
For any index $j\in\nat^{+}$, we define $2\Sigma_j\auxpdatisp(t(n),s(n))$ to be  the collection of all decision problems solvable by \emph{two-way alternating auxiliary pushdown automata} running within time $t(n)$ using space at most $s(n)$ with at most $j$ alternations starting with existential inner states. When $j=1$, such machines are succinctly called \emph{2aux-npda's}, which will be used in the proof of Proposition \ref{counter-reduction}.
It is known that $2\Sigma_1\auxpdatisp(n^{O(1)},O(\log{n}))$ coincides with $\logcfl$ \cite{Sud78} and $2\Sigma_2\auxpdatisp(n^{O(1)},O(\log{n}))$ coincides with $\np$ \cite{JK89}.
Moreover, it is possible to simulate the polynomial time-bounded behaviors of $k$ counters using an $O(\log{n})$-space bounded auxiliary work tape. From these facts, the lemma follows immediately.
\end{proofsketch}

\begin{proposition}\label{counter-reduction}
$\nl=\mathrm{2N4CA}$ and $\logcfl = \mathrm{2NPD3CA}$.
\end{proposition}

\begin{proofsketch}
Following an argument of Minsky \cite{Min67}, given a log-space NTM (or a log-space 2aux-npda), we first simulate the behavior of an $O(\log{n})$-space auxiliary work tape by two stacks whose alphabets are of the form $\{0,1,\bot\}$. Since the heights of such stacks are upper-bounded by $O(\log{n})$, the stacks can be further simulated by multiple counters whose heights are  all $n^{O(1)}$-bounded.
Let $M$ denote the obtained 2ncta (or 2npdcta) running in polynomial time.

We remark that Minsky's method of reducing the number of counters down to two does not work for machines with few computational resources. Thus, we need to seek other methods. For this purpose, we review the result of \cite{Yam23a}, which makes it possible to simulate two counters by one counter with the heavy use of an appropriately defined ``pairing''  function.

\begin{claim}\cite{Yam23a}\label{claim-reduction}
There exists a fixed deterministic procedure by which any single move of push/pop operations on two counters of $M$ can be simulated by a series of $n^{O(1)}$ push/pop operations on one counter with the help of 3 extra counters. These 3 extra counters are emptied after each simulation and thus they are reusable for any other purposes. If a stack is further available, then one extra counter can be simulated by this stack.
\end{claim}

The recursive applications of this simulation procedure eventually reduce the number of counters down to four.
If we freely use a stack during the procedure, then we can further reduce the number of counters down to three.

The first part of the proposition then follows by combining the above claim  with Lemma \ref{counter-inclusion}. If we use one counter as a stack, then we can further reduce 4 counters to 3 counters. This proves the second part of the proposition.
\end{proofsketch}

\section{Combinatorial Number Problems}\label{sec:number-theory}

We study the computational complexity of decision problems whose input instances are composed of unarized positive integers.

\subsection{Variations of the Unary 0-1 Knapsack Problem}

The starting point of our study on the computational analyses of decision problems with unarized integer inputs is the \emph{unary 0-1 knapsack problem}, which was introduced in 1985 by Cook \cite{Coo85} as a unary analogue of the  \emph{knapsack problem}.

\ms
{\sc Unary 0-1 Knapsack Problem}  ({\sc UK}):
\renewcommand{\labelitemi}{$\circ$}
\begin{itemize}\vs{-1}
  \setlength{\topsep}{-2mm}%
  \setlength{\itemsep}{1mm}%
  \setlength{\parskip}{0cm}%

\item {\sc Instance:} $(1^{b},1^{a_1},1^{a_2},\ldots,1^{a_n})$, where $n\in\nat^{+}$ and $b,a_1,a_2,\ldots,a_n$ are all positive integers.

\item {\sc Question:} is there a subset $S$ of $[n]$ satisfying  $\sum_{i\in S}a_i = b$?
\end{itemize}

It seems more natural to view $\mathrm{UK}$ as a unary analogue of the \emph{subset sum problem}, which is closely related to the \emph{2-partition problem}. Let us consider a unary analogue of this 2-partition problem.

\ms
{\sc Unary 2 Partition Problem}  ({\sc U2PART}):
\renewcommand{\labelitemi}{$\circ$}
\begin{itemize}\vs{-1}
  \setlength{\topsep}{-2mm}%
  \setlength{\itemsep}{1mm}%
  \setlength{\parskip}{0cm}%

\item {\sc Instance:} $(1^{a_1},1^{a_2},\ldots,1^{a_n})$, where $n\in\nat^{+}$ and $a_1,a_2,\ldots,a_n$ are all positive integers.

\item {\sc Question:} is there a subset $S$ of $[n]$ such that $\sum_{i\in S}a_i = \sum_{i\in \overline{S}}a_i$, where $\overline{S}= [n]-S$?
\end{itemize}


The original knapsack problem and the subset sum problem are both proven  in 1972 by Karp \cite{Kar72} to be $\np$-complete.
The problems $\mathrm{UK}$ and $\mathrm{U2PART}$, in contrast, situated within $\nl$.

\begin{lemma}\label{UK-upper-bound}
(1) $\mathrm{UK}$ is in $\mathrm{1t1NCA}$. (2) $\mathrm{U2PART}$ is in both $\mathrm{1NCA}$ and $\mathrm{1t1N2CA}$.
\end{lemma}

We further introduce two variants of $\mathrm{UK}$ and $\mathrm{U2PART}$, called $\mathrm{AmbUK}$ and $\mathrm{UASubSum}$ as follows.

\ms
{\sc Ambiguous UK Problem}  (AmbUK):
\renewcommand{\labelitemi}{$\circ$}
\begin{itemize}\vs{-1}
  \setlength{\topsep}{-2mm}%
  \setlength{\itemsep}{1mm}%
  \setlength{\parskip}{0cm}%
\item {\sc Instance:} $((1^{b_1},1^{b_2},\ldots, 1^{b_m}),(1^{a_1},1^{a_2},\ldots,1^{a_n}))$, where $n,m\in\nat^{+}$ and $b_1,b_2,\ldots,b_m, a_1,a_2,\ldots,a_n$ are all positive integers.

\item {\sc Question:} are there an index $j\in[m]$ and a subset $S$ of $[n]$ satisfying $b_j =  \sum_{i\in S}a_i$?
\end{itemize}

\ms
{\sc Unary Approximate Subset Sum Problem}  (UASubSum):
\renewcommand{\labelitemi}{$\circ$}
\begin{itemize}\vs{-1}
  \setlength{\topsep}{-2mm}%
  \setlength{\itemsep}{1mm}%
  \setlength{\parskip}{0cm}%

\item {\sc Instance:} $((1^{b_1},1^{b_2}),(1^{a_1},1^{a_2},\ldots,1^{a_n}))$, where $n\in\nat^{+}$ and $b_1,b_2,a_1,a_2,\ldots,a_n$ are positive integers.

\item {\sc Question:} is there a subset $S$ of $[n]$ such that $b_1\leq \sum_{i\in S}a_i \leq b_2$?
\end{itemize}

\begin{lemma}\label{AmbUK-in-1t1NCA}
(1) $\mathrm{AmbUK}$ is in $\mathrm{1t1NCA}$. (2) $\mathrm{UASubSum}$ is in $\mathrm{1t1N2CA}$.
\end{lemma}

Under two different types of log-space reductions, the computational complexities of $\mathrm{U2PART}$, $\mathrm{AmbUK}$, and  $\mathrm{UASubSum}$ are all equal to that of $\mathrm{UK}$.

\begin{proposition}\label{UK-variants-reduction}
(1) $\mathrm{UK} \equiv^{\mathrm{L}}_{m} \mathrm{U2PART}$. (2)
$\mathrm{UK} \equiv^{\mathrm{L}}_{tt} \mathrm{AmbUK}$.  (3)
$\mathrm{UK}  \equiv^{\mathrm{L}}_{m} \mathrm{UASubSum}$.
\end{proposition}

\begin{proofsketch}
(1) We begin with showing that $\mathrm{UK} \equiv^{\mathrm{L}}_{m} \mathrm{U2PART}$. Our first goal is to prove that $\mathrm{UK} \leq^{\mathrm{L}}_{m} \mathrm{U2PART}$ by constructing an $\dl$-m-reduction function $f$.
Let $x=(1^b,1^{a_1},\ldots,1^{a_n})$ and let $c=\sum_{i\in[n]}a_i$. (a) If $b=c/2$, then we set $f(x)=x$. (b) If $b<c/2$, then we add a new entry $a_{n+1}=c-2b$ and set $f(x)=(1^{a_1},\ldots,1^{a_n},1^{a_{n+1}})$. (c) If $b>c/2$, then we exchange between $b$ and $c-b$ and apply (b).
Our second goal is to show that $\mathrm{U2PART} \leq^{\mathrm{L}}_{m} \mathrm{UK}$. Letting $c=\sum_{i}a_i$, if $c$ is odd, then we set $f(x)$ to be any fixed rejected input. Otherwise, we set $b=c/2$ and define $f(x)=(1^{b},1^{a_1},\ldots,1^{a_n})$.

(2) It is obvious that $\mathrm{UK} \leq^{\mathrm{L}}_{m} \mathrm{AmbUK}$ by setting $m=1$ in the definition of $\mathrm{AmbUK}$. Next, we show that $\mathrm{AmbUK} \leq^{\mathrm{L}}_{tt} \mathrm{UK}$. Let $w= ((1^{b_1},1^{b_2},\ldots,1^{b_m}),(1^{a_1},1^{a_2},\ldots,1^{a_n}))$ be any instance of $\mathrm{AmbUK}$. We define an $\dl$-tt-reduction function $f$ as $f(w) = (f_1(w),f_2(w),\ldots,f_m(w))$, where $f_i(w) = (1^{b_{i}},1^{a_1},\ldots,1^{a_n})$. Clearly, $f$ belongs to $\fl$. We also define a truth table $E$ as $E(e_1,e_2,\ldots,e_m) = \bigvee_{i=1}^{m}e_i$. It then follows that $w\in \mathrm{AmbUK}$ iff there exists an index $i\in[m]$ for which $f_i(w)\in\mathrm{UK}$ iff $E(f_1(w),f_2(w),\ldots,f_m(w))=1$. Therefore, $\mathrm{AmbUK} \leq^{\mathrm{L}}_{tt} \mathrm{UK}$ follows.

(3) To show that $\mathrm{UK} \leq^{\mathrm{L}}_{m} \mathrm{UASubSum}$, it suffices to set $b_1=b_2=b$ and define the desired reduction function $f$ to map $(1^b,1^{a_1},\ldots,1^{a_n})$ to $((1^{b_1},1^{b_2}),(1^{a_1},\ldots,1^{a_n}))$.
On the contrary, we wish to verify that $\mathrm{UASubSum} \leq^{\mathrm{L}}_{m} \mathrm{UK}$.
Let $w=((1^{b_1},1^{b_2}), (1^{a_1},\ldots,1^{a_n}))$ be any input to $\mathrm{UASubSum}$.
If $b_1=b_2$, then the reduction is trivial.
In the case of $b_1>b_2$, it suffices to define $f(w)=(1,1^2)$ since $w\notin \mathrm{UASubSum}$. In what follows, we assume that $b_1<b_2$.
For any $S\subseteq[n]$, we define $a_{S}=\sum_{i\in S}a_i$. If $a_{[n]}<b_1$, then we construct $f(w)=(1,1^2)$, which is obviously not in $\mathrm{UASubSum}$. If $b_1\leq a_{[n]}\leq b_2$, then we define $x=(1,1)$. Next, we assume that $b_2<a_{[n]}$.
Let us define $a_{n+i} =a_{[n]}+i-1$ for any $i\in[b_2-b_1+1]$ and let  $b_{max}=b_2+a_{[n]}$. We then set $f(w)=(1^{b_{max}},1^{a_1},1^{a_2},\ldots, 1^{a_n},1^{a_{n+1}},1^{a_{n+2}},\ldots, 1^{a_{n+b_2-b_1+1}})$.
This concludes that $w\in \mathrm{UASubSum}$ iff $f(w)\in \mathrm{UK}$.
\end{proofsketch}


Jenner \cite{Jen95} studied a variant of $\mathrm{UK}$, which we intend to call by the \emph{shift-unary 0-1 knapsack problem} (shift-UK) in order to signify the use of the shift-unary representation. She proved that this problem is actually $\nl$-complete.

\ms
{\sc Shift-Unary 0-1 Knapsack Problem}  (shift-UK):
\renewcommand{\labelitemi}{$\circ$}
\begin{itemize}\vs{-1}
  \setlength{\topsep}{-2mm}%
  \setlength{\itemsep}{1mm}%
  \setlength{\parskip}{0cm}%

\item {\sc Instance:} $[1^q,1^s]$ and a series $([1^{p_1},1^{t_1}], [1^{p_2},1^{t_2}],\ldots,[1^{p_n},1^{t_n}])$ of nonnegative integers represented in shift-unary, where $q,p_1,p_2,\ldots,p_n$ are all positive integers and $s,t_1,t_2,\ldots,t_n$ are all nonnegative integers.

\item {\sc Question:} is there a subset $S$ of $[n]$ such that $\sum_{i\in S}p_i 2^{t_i} = q 2^s$?
\end{itemize}

In a similar way, we can define the shift-unary representation versions of $\mathrm{AmbUK}$, $\mathrm{UASubSum}$, and $\mathrm{U2PART}$, denoted by
$\shift\mathrm{AmbUK}$, $\shift\mathrm{UASubSum}$, and  $\shift\mathrm{U2PART}$, respectively.

\begin{lemma}\label{shift-UP-bound}
The problem $\shift\mathrm{UK}$ is in $\mathrm{1N6CA}$.
\end{lemma}

\begin{proposition}\label{shift-UK-equivalence}
$\shift\mathrm{UK} \equiv^{\mathrm{L}}_{m}  \shift\mathrm{U2PART} \equiv^{\mathrm{L}}_{m}
\shift\mathrm{AmbUK} \equiv^{\mathrm{L}}_{m}
\shift\mathrm{UASubSum}$.
\end{proposition}

\begin{proofsketch}
We abbreviate the set $\{\shift\mathrm{U2PART}, \shift\mathrm{AmbUK}, \shift\mathrm{UASubSum} \}$ as $\SSS$.
It is easy to obtain, by modifying the proof of Proposition \ref{UK-variants-reduction}, that
$\shift\mathrm{UK}\leq^{\mathrm{L}}_{m}\CC$ for any $\CC\in\SSS$.
To show that $\CC \leq^{\mathrm{L}}_{m} \shift\mathrm{UK}$ for any $\CC\in\SSS$, it suffices to show that $\CC$ belongs to $\nl$ because the $\dl$-m-completeness of $\shift\mathrm{UK}$ \cite{Jen95} guarantees that  $\CC\leq^{\mathrm{L}}_{m}\shift\mathrm{UK}$.
In the case of $\CC = \shift\mathrm{UASubSum}$,
we consider the following algorithm. On input of the form $(([1^{q_1},1^{s_1}], [1^{q_2},1^{s_2}]), ([1^{p_1},1^{t_1}], \ldots, [1^{p_n},1^{t_n}]))$, we nondeterministically choose a number $k\in[t]$ and indices $i_1,i_2,\ldots,i_k\in[n]$ and check that $q_12^{s_1}\leq \sum_{i\in S} p_i2^{t_i} \leq q_22^{s_2}$.
Note that, by Lemma \ref{function-sum}, the sum $\sum_{i\in S} p_i2^{t_i}$ can be computed using only log space. Hence, $C$ is in $\nl$.
The other cases can be similarly handled.
\end{proofsketch}

From Proposition \ref{shift-UK-equivalence}, since $\shift\mathrm{UK}$ is $\nl$-complete under $\dl$-m-reductions \cite{Jen95}, we immediately obtain the following corollary.

\begin{corollary}
The problems $\shift\mathrm{U2PART}$, $\shift\mathrm{AmbUK}$, and $\shift\mathrm{UASubSum}$ are all $\nl$-complete.
\end{corollary}


For later references, we introduce another variant of $\mathrm{UK}$. This  variant requires a prohibition of certain successive choices of unarized  integer inputs.

\ms
{\sc UK with Exception}  (UKEXC):
\renewcommand{\labelitemi}{$\circ$}
\begin{itemize}\vs{-1}
  \setlength{\topsep}{-2mm}%
  \setlength{\itemsep}{1mm}%
  \setlength{\parskip}{0cm}%

\item {\sc Instance:} $(1^{b},1^{a_1},1^{a_2},\ldots,1^{a_n})$ and $EXC\subseteq \{(i,j) \mid i,j\in[n], i<j\}$ given as a set of pairs $(1^{i},1^{j})$, where $n\in\nat^{+}$ and $b,a_1,a_2,\ldots,a_n$ are in $\nat^{+}$.

\item {\sc Question:} is there a subset $S=\{i_1,i_2,\ldots,i_k\}$ of  $[n]$ with $k\in\nat^{+}$ and $i_1<i_2<\cdots <i_k$ such that (i) $\sum_{i\in S}a_i = b$ and (ii) no $j\in[k-1]$ satisfies $(i_j,i_{j+1})\in EXC$?
\end{itemize}

Note that, when $EXC=\setempty$, $\mathrm{UKEXC}$ is equivalent to $\mathrm{UK}$.

\begin{lemma}
$\mathrm{UKEXC}$ is in $\mathrm{2N3CA}$.
\end{lemma}

\begin{proofsketch}
We nondeterministically choose $1^{a_i}$ by reading an input from left to right. We use two counters to remember the index $i$ (in the form of $1^{i}$) for checking that the next possible choice, say, $1^{a_j}$ satisfies $(i,j)\notin EXC$. The third counter is used to store $1^{b}$ and then sequentially pop $1^{a_i}$ for the chosen indices $i$.
\end{proofsketch}

We also introduce another variant of $\mathrm{UK}$, which concerns  simultaneous handling of input integers.

\ms
{\sc Simultaneous Unary 0-1 Knapsack Problem}  (SUK):
\renewcommand{\labelitemi}{$\circ$}
\begin{itemize}\vs{-1}
  \setlength{\topsep}{-2mm}%
  \setlength{\itemsep}{1mm}%
  \setlength{\parskip}{0cm}%

\item {\sc Instance:} $(1^{b_1},1^{a_{11}}, 1^{a_{12}},\ldots,1^{a_{1n}}), \ldots,$ $(1^{b_m},1^{a_{m1}},1^{a_{m2}}, \ldots,1^{a_{mn}})$, where $m,n\in\nat^{+}$ and $b_i$ and $a_{ij}$ ($i\in[n],j\in[m]$) are all positive integers.

\item {\sc Question:} is there a subset $S\subseteq [n]$ such that  $\sum_{j\in S}a_{ij} = b_i$ for all indices $i\in[m]$?
\end{itemize}

The complexity class $\mathrm{1t}1\Sigma_2\mathrm{PDCA}$ is the one-way restriction of $\mathrm{1t2}\Sigma_2\mathrm{PDCA}$.

\begin{lemma}
The problem $\mathrm{SUK}$ is in $\mathrm{1t}1\Sigma_2\mathrm{PDCA}$.
\end{lemma}

\begin{proofsketch}
To recognize $\mathrm{SUK}$, let us consider a \emph{one-way alternating counter pushdown automaton} (or a 1apdcta) that behaves as follows. Given an input $x$ of the form $(1^{b_1},1^{a_{11}}, 1^{a_{12}},\ldots,1^{a_{1n}}), \ldots,$ $(1^{b_m},1^{a_{m1}},1^{a_{m2}}, \ldots,1^{a_{mn}})$, we call each segment $(1^{b_i},1^{a_{i1}}, 1^{a_{i2}},\ldots,1^{a_{in}})$ of $x$ by the \emph{$i$th block} of $x$.

In existential inner states, we first choose a string $w\in\{0,1\}^*$ and push it into a stack, where $w=e_1e_2\cdots e_n$ indicates that, for any $j\in[n]$,  we select the $j$th entry from every block exactly when $e_j=1$. Let $A_w=\{j\in[n]\mid e_j=1\}$. In universal inner states, we then check whether $b_i=\sum_{j\in A_w}a_{ij}$ holds for all $i\in[m]$.
This last procedure is achieved by first storing $1^{b_i}$ into a counter.
As popping the values $e_j$ one by one from the stack,  if $j\in A_w$, then we decrease the counter by $a_{ij}$. Otherwise, we do nothing. When either the stack gets empty or the assigned block $(1^{a_{i1}},\ldots,1^{a_{in}})$ of $x$ is finished, if the stack is empty and the counter becomes $0$, then we accept $x$; otherwise, we reject $x$. Note that the stack and the counter make only 1-turns and the input-tape head moves only in one direction.
\end{proofsketch}

\subsection{Unary Bounded Correspondence Problem}

We turn our attention to the \emph{bounded Post correspondence problem} (BPCP), which is a well-known problem of determining whether, given a set $\{(a_i,b_i)\}_{i\in[n]}$ of binary string pairs and a number $k\geq1$, a certain sequence $(i_1,i_2,\ldots,i_t)$ of elements in $[n]$ with $t\leq k$ satisfies $a_{i_1}a_{i_2}\cdots a_{i_t} = b_{i_1}b_{i_2}\cdots b_{i_t}$. This problem is known to be $\np$-complete \cite{CHS74}.
When we  replace binary strings $a_i$ and $b_i$ by unary strings, we obtain the following ``unary'' variant of $\mathrm{PCP}$.

\ms
{\sc Unary Bounded Correspondence Problem}  ({\sc UBCP}):
\renewcommand{\labelitemi}{$\circ$}
\begin{itemize}\vs{-1}
  \setlength{\topsep}{-2mm}%
  \setlength{\itemsep}{1mm}%
  \setlength{\parskip}{0cm}%

\item {\sc Instance:} $((a_1,b_1),(a_2,b_2), \ldots,(a_n,b_n))$ for unary strings $a_1,a_2,\ldots,a_n,b_1,b_2,\ldots,b_n\in 1^{+}$ and $1^k$ for $k\in\nat^{+}$.

\item {\sc Question:} is there a sequence  $(i_1,i_2,\ldots,i_t)$ of elements in $[n]$ with $t\in[k]$ satisfying $a_{i_1}a_{i_2}\cdots a_{i_t} = b_{i_1}b_{i_2}\cdots b_{i_t}$?
\end{itemize}

Since $a_i$ and $b_i$ are unary strings, the above requirement $a_{i_1}\cdots a_{i_t}=b_{i_1}\cdots b_{i_t}$ is equivalent to $\sum_{j\in S}|a_j| = \sum_{j\in S}|b_j|$, where $|a_j|$ and $|b_j|$ mean the lengths of strings $a_j$ and $b_j$, respectively.

\begin{lemma}\label{UBCP-vs-1t1N2CA}
$\mathrm{UBCP}$ belongs to $\mathrm{1t1N2CA}$.
\end{lemma}

\begin{proposition}\label{UK-vs-UBCP}
$\mathrm{UK}\leq^{\mathrm{L}}_{m} \mathrm{UBCP} \leq^{\mathrm{L}}_{m} \mathrm{AmbUK}$.
\end{proposition}

\begin{proofsketch}
Here, we only show the last reduction  $\mathrm{UBCP}\leq^{\mathrm{L}}_{m}\mathrm{AmbUK}$. Let $w = ((a_1,b_1),\ldots,(a_n,b_n))$ be any instance to $\mathrm{UBCP}$.
Assume that, for any $i\in[n]$, $a_i=1^{k_i}$ and $b_i=1^{l_i}$ for certain numbers $k_i,l_i\in\nat^{+}$. Let $m=\max\{\sum_{i\in[n]}k_i, \sum_{i\in[n]}l_i\}$.  Note that $|S|\leq \sum_{i\in S}k_i\leq m$ and $|S|\leq \sum_{i\in S}l_i\leq m$ for any $S\subseteq [n]$. We then set $c_i=m-(k_i-l_i)$ for each $i\in[n]$. We remark that $\sum_{i\in S}c_i = m|S|-(\sum_{i\in S}k_i - \sum_{i\in S}l_i) \geq m|S|-(m-|S|) = m(|S|-1)+|S|\geq0$ for any nonempty subset $S$ of $[n]$. If $w\in \mathrm{UBCP}$, then there is a nonempty set $S\subseteq[n]$ such that $\sum_{i\in S}k_i = \sum_{i\in S}l_i$, that is, $\sum_{i\in S}(k_i-l_i)=0$. It then follows that $\sum_{i\in S}c_i = m|S|$.

We also check if $\sum_{i\in[n]}k_i = \sum_{i\in [n]}l_i$ or if $k_{i_0}=l_{i_0}$ for a certain index $i_0\in[n]$. If this is true, then we know that $S=[n]$ or $\{i_0\}$. Otherwise, we set $d_i= i\cdot m$ for each $i\in[2,n-1]_{\integer}$ and then define $u = ((1^{d_2},1^{d_3},\ldots,1^{d_{n-1}}), (1^{c_1},1^{c_2},\ldots, 1^{c_n}))$.
It follows that $w\in \mathrm{UBCP}$ iff $u\in\mathrm{AmbUK}$.
\end{proofsketch}

\begin{corollary}
$\mathrm{UK}\equiv^{\mathrm{L}}_{tt} \mathrm{UBCP}$.
\end{corollary}

\section{Graph Problems}\label{sec:graph-related}

We look into decision problems that are related to graphs, in particular, weighted graphs in which either vertices or edges are labeled with ``weights''. Here, we study the computational complexity of these specific problems.


We begin with edge-weighted path problems, in which we search for a simple path whose weight matches a target unarized number, which is given in unary.
We consider, in particular, \emph{directed acyclic graphs} (or dags) whose edges are further labeled with weights, which are given in unary.

For the purpose of this work, a dag $G=(V,E)$ given as part of inputs to an underlying machine is assumed to satisfy that
$V$ is a subset of $\nat^{+}$, e.g., $V=\{i_1,i_2,\ldots,i_n\}$ for $i_1,i_2,\ldots,i_n\in\nat^{+}$.
We also use the following specific encoding of $G$. Given a vertex $i$, we write  $E[i]$ for the set of all adjacent vertices entering from $i$, namely, $\{j\in V\mid (i,j)\in E\}$.  When $E[i]$ equals $\{j_1,j_2,\ldots,j_n\}$ with $j_1<j_2<\cdots <j_n$, we express it in the ``multi-unary'' form  of $(1^i,1^{j_1},1^{j_2},\ldots,1^{j_n})$. The \emph{unary adjacency list representation} $UAL_G$ of $G$ is of the form $((1^{i_1},1^{j_{i_1,1}},1^{j_{i_1,2}},\ldots,1^{j_{i_1,k_1}}),\ldots, (1^{i_n},1^{j_{i_n,1}},1^{j_{i_n,2}},\ldots,1^{j_{i_n,k_n}}))$ with $k_i\in\nat$ if $V=\{i_1,i_2,\ldots,i_n\}$ with $i_1<i_2<\cdots <i_n$ and $E[i_s] = \{j_{i_s,1},j_{i_s,2},\ldots,j_{i_s,k_s}\}$ with $j_{i_s,1}<j_{i_s,2}<\cdots <j_{i_s,k_s}$ for any $s\in[n]$.

\ms
{\sc  Edge-Weighted Path Problem} ({\sc EWPP})
\renewcommand{\labelitemi}{$\circ$}
\begin{itemize}\vs{-1}
  \setlength{\topsep}{-2mm}%
  \setlength{\itemsep}{1mm}%
  \setlength{\parskip}{0cm}%

\item {\sc Instance:} a dag $G=(V,E)$ with $V\subseteq \nat^{+}$ given in the form of $UAL_{G}$, a vertex $s\in V$ given as $1^s$, edge weights $w(i,j)\in\nat^{+}$ given as  $1^{w(i,j)}$ for all edges $(i,j)\in E$, and $1^c$ with $c\in\nat^{+}$, provided that edges are enumerated according to the linear order $<_{[n]}$.

\item {\sc Question:} is there  a vertex $v\in V$ such that the total edge weight of a path from $s$ to $v$ equals $c$?
\end{itemize}
\ms

In comparison, the \emph{graph connectivity problem} for directed ``unweighted'' graphs (DSTCON) is known to be $\nl$-complete \cite{JLL76}.

\begin{lemma}\label{EWPP-in-NL}
$\mathrm{EWPP}$ is in $\nl$.
\end{lemma}

When each edge weight is $1$, the total weight of a path is the same as the length of the path. This fact makes us introduce another decision problem. A \emph{sink} of a directed graph is a vertex of outdegree $0$ in the graph.

\ms
{\sc  Exact Path Length Problem} ({\sc EPLP})
\renewcommand{\labelitemi}{$\circ$}
\begin{itemize}\vs{-1}
  \setlength{\topsep}{-2mm}%
  \setlength{\itemsep}{1mm}%
  \setlength{\parskip}{0cm}%

\item {\sc Instance:} a dag $G=(V,E)$ with $V\subseteq \nat^{+}$ given as $UAL_{G}$, a vertex $s\in V$ given as $1^s$, and $1^c$ with $c\in\nat^{+}$.

\item {\sc Question:} is there a sink $v$ of $G$ such that a path from $s$ to $v$ has length exactly $c$?
\end{itemize}

\begin{proposition}\label{EWPP-EPLP}
$\mathrm{EWPP} \equiv^{\mathrm{L}}_{m} \mathrm{EPLP}$.
\end{proposition}

\begin{proofsketch}
Firstly, we show that $\mathrm{EPLP}\Lmreduces \mathrm{EWPP}$. Given an instance $x=(G,s,1^c)$ of $\mathrm{EPLP}$ with $G=(V,E)$, we define another instance, say, $y$ of $\mathrm{EWPP}$ as follows.
For every leaf $v$ of $G$, we add a new vertex $\bar{v}$ and a new edge $(v,\bar{v})$ to $G$ and obtain the new vertex set $V'$ and the new edge set $E'$. Consider the resulting graph $G'=(V',E')$. Let $n=|V'|$. We define $w(u,v)=1$ for all pairs $(u,v)\in E$ and additionally we set $w(v,\bar{v})=n+1$ for any old leaf $v\in V$. Let $c'=c+n+1$.
We define $y=(G',\{w(e)\}_{e\in E'},1^{c'})$. We want to verify that (*) $x\in \mathrm{EPLP}$ iff $y\in \mathrm{EWPP}$.

To prove (*), let us assume that $x\in\mathrm{EPLP}$. We then take a length-$c$ path $\gamma=(v_0,v_1,\ldots,v_c)$ of $G$ with $v_0=s$ and $v_c$ is a leaf. Consider $y$ and $\gamma^{(+)}=(v_0,v_1,\ldots,v_c,\bar{v}_c)$. The total weight of $\gamma^{(+)}$ in $G'$ is $c+w(v_c,\bar{v}_c) = c+n+1 =c'$. Hence, $y$ is in $\mathrm{EWPP}$.
Conversely, assume that $y\in \mathrm{EWPP}$ and consider a path $\gamma=(v_0,v_1,\ldots,v_k)$ of $G'$ with $v_0=s$ satisfying $\sum_{i=0}^{k}w(v_i,v_{i+1}) =c'$. Since $n=|V'|$, $\gamma$ must include a leaf. Hence, $v_k$ must be $\bar{v}_{k-1}$. Moreover, $k-1=c$ follows. The path $\gamma^{(-)}=(v_0,v_1,\ldots,v_{k-1})$ is a path from $s$ to a leaf of $G$ and its length is exactly $c$. This implies that $x\in\mathrm{EPLP}$.

Secondly, we show that $\mathrm{EWPP}\Lmreduces \mathrm{EPLP}$. Let $x=(G,s,\{w(e)\}_{e\in E},1^c)$ be any instance of $\mathrm{EWPP}$. We construct an instance of $\mathrm{EPLP}$ as follows. For each $(u,v)\in E$ with $w(u,v)=1^k$,
we introduce $k+1$ extra vertices $\{\bar{v},u'_1,u'_2,\ldots,u'_{k}\}$ and extra edges $\{(u,u'_1),(u'_1,u'_2), \ldots,(u'_{i},u'_{i+1}), (u'_{k},v),(u'_{k},\bar{v}) \mid i\in[k-2]\}$.  Those edges form two paths from $u$ to $v$ and $u$ to $\bar{v}$ of length exactly  $k$.
Notice that $\bar{v}$ becomes a new leaf.
Let $V'$ and $E'$ denote the extended sets of $V$ and $E$, respectively, and set $G'=(V',E')$. For the instance $y=(G',s,1^c)$, it is possible to prove  that $x\in \mathrm{EWPP}$ iff $y\in \mathrm{EPLP}$.
\end{proofsketch}

We then obtain the following $\nl$-completeness result.

\begin{proposition}\label{EPLP-L-complete}
$\mathrm{EWPP}$ and $\mathrm{EPLP}$ are  both $\dl$-m-complete for $\nl$.
\end{proposition}

\begin{proofsketch}
Here, we wish to prove (*) $L\leq^{\mathrm{L}}_{m}\mathrm{EPLP}$ for any language $L$ in $\mathrm{1NCA}$. If this is true, then all languages in $\nl$ are $\dl$-m-reducible to $\mathrm{EPLP}$ since $\mathrm{LOG(1NCA)}=\nl$.
Moreover, since $\mathrm{EWPP}$ belongs to $\nl$ by Lemma \ref{EWPP-in-NL}, $\mathrm{EPLP}$ is also in $\nl$.
Therefore, $\mathrm{EPLP}$ is $\dl$-m-complete for $\nl$.
Since $\mathrm{EWPP}\equiv^{\dl}_{m} \mathrm{EPLP}$ by Proposition  \ref{EWPP-EPLP}, we also obtain the $\nl$-completeness of $\mathrm{EWPP}$.

To show the statement (*), let us take any 1ncta $M$ and consider surface configurations of $M$ on input $x$.
Note that, since surface configurations have size $O(\log{|x|})$, we can express them as unary-form positive integers.
Between two surface configurations, we define a single-step transition relation $\vdash_{M}$.
We then define a computation graph $G_x=(V_x,E_x)$ of $M$ on the input $x$. The vertex set $V_x$ is composed of all possible surface configurations of $M$ on $x$. We further define $E_x$ to be the set of all pairs $(v_1,v_2)$ of surface configurations satisfying $v_1\vdash_{M} v_2$. The root $s$ of $G_x$ is set to be the initial surface configuration of $M$.
It then follows that $x\in L(M)$ iff $(UAL_{G_x},1^s,1^{|x|+2})\in\mathrm{EPLP}$.
\end{proofsketch}


Let us  argue how $\mathrm{EWPP}$ is related to $\mathrm{UKEXC}$ and $\mathrm{UK}$. To see the desired relationships, we introduce two restricted variants of $\mathrm{EWPP}$. We say that a dag $G=(V,E)$ with $V\subseteq \nat^{+}$ is \emph{topologically ordered} if, for any two vertices $i,j\in V$, $(i,j)\in E$ implies $i<j$. We define $\mathrm{TO\mbox{-}EWPP}$ as the restriction of $\mathrm{EWPP}$ onto instances that are topologically ordered. 
A dag $G=(V,E)$ is \emph{edge-closed} if, for any three vertices $u,v,w\in V$, (1) $(u,v)\in E$ and $(v,w)\in E$ imply $(u,w)\in E$ and (2) $(u,w)\in E$ and $(v,w)\in E$ imply either $(u,v)\in E$ or $(v,u)\in E$. We write $\mathrm{EC\mbox{-}EWPP}$ for $\mathrm{TO\mbox{-}EWPP}$ whose instance graphs are all edge-closed.

\begin{proposition}\label{EWPP-and-UK}
(1) $\mathrm{TO\mbox{-}EWPP} \equiv^{\mathrm{L}}_{m} \mathrm{UKEXC}$.
(2) $\mathrm{EC\mbox{-}EWPP}\equiv^{\mathrm{L}}_{m} \mathrm{UK}$.
\end{proposition}

\begin{proofsketch}
In what follows, we focus only on (1). We first show that $\mathrm{UKEXC}\Lmreduces \mathrm{TO\mbox{-}EWPP}$.
Given an instance $x= ((1^b,1^{a_1},\ldots,1^{a_n}), EXC)$ of $\mathrm{UKEXC}$, we construct an associated instance, say, $y$ of $\mathrm{TO\mbox{-}EWPP}$ as follows.
We then define $V=\{s,v_1,v_2,\ldots,v_n\}$ and $E=\{(v_i,v_j) \mid i,j\in[n], i<j, (i,j)\notin EXC\} \cup\{(s,v_j)\mid j\in[n]\}$. It follows that $G=(V,E)$ is directed, acyclic, and topologically ordered.
Furthermore, we define edge weights as $w(v_i,v_j) =a_i$ and $w(s,v_j)=1$ for any $(v_i,v_j),(s,v_j)\in E$. Finally, we set $c=b+1$.
Let us consider $y=(G,\{1^{w(e)}\}_{e\in E},1^c)$.
It is possible for us to verify
that $x\in \mathrm{UKEXC}$ iff $y\in \mathrm{TO\mbox{-}EWPP}$. This implies
$\mathrm{UKEXC}\Lmreduces \mathrm{TO\mbox{-}EWPP}$.

Conversely, we intend to verify that $\mathrm{TO\mbox{-}EWPP}\Lmreduces \mathrm{UKEXC}$. Assume that $x=(G,s,\{1^{w(e)}\}_{e\in E},1^c)$ with $G=(V,E)$ is any instance of $\mathrm{TO\mbox{-}EWPP}$. An associated instance $y$ of $\mathrm{UKEXC}$ is constructed as follows.
Firstly, we assume that $G$ is acyclic because, otherwise, we convert $G$ to an acyclic graph $G'=(V',E')$ by setting $V'=\{((i,v)\mid i\in[|V|],v\in V\}$ and $E'=\{((i,u),(i+1,v)) \mid i\in[|V|-1], (u,v)\in E\}$.
For simplicity, we further assume that $V=[2,n]_{\integer}$ and $s$ is vertex $2$.
We define an encoding $\pair{i,j}$ of two distinct vertices $i,j\in[n]$ as $\pair{i,j}=(i-1)n+j$.
We consider only a unique connected component including $s$ because any other connected component is irrelevant. Hereafter, we assume that this  connected component contains all vertices in $V$. We expand $G$ to $G'=(V',E')$ by setting $V'=V\cup\{1\}$ and $E'=E\cup\{(1,j)\mid j\in[2,n]_{\integer}\}$. As for new weights, we set $w'(i,j) = w(i,j)$ for any $(i,j)\in E$ and $w'(1,j)=w_*+1$ for any $j\in[2,n]_{\integer}$, where $w_{*}=\sum_{(i,j)\in E}w(i,j)$.

We further set $a_{\pair{i,j}}=w(i,j)$ if $(i,j)\in E$.
Moreover, let $a_{\pair{1,j}}=w_{*}+1$ for any $j\in[2,n]_{\integer}$. For any other pair $(i,j)$, the value $a_{\pair{i,j}}$ is not defined.
Let $A$ denote the set of all $a_{\pair{i,j}}$'s that are defined. Assume that $A$ has the form $\{a_{k_1},a_{k_2},\ldots,a_{k_m}\}$ with $m\leq n^2$, where  $k_1<k_2<\cdots <k_m$. We write $K$ for the set $\{k_1,k_2,\ldots,k_m\}$.
The set $EXC$ is defined to be the union of the following sets: $\{(\pair{i,j},\pair{i',j'})\mid i,j,i',j'\in [n], (\pair{i,j} \not< \pair{i',j'}\vee j\neq i'\vee i=j'), (i,v)\in E, ({i'},{j'})\in E\}$, $\{(\pair{i,j},\pair{i',j'}) \mid i,i',j,j'\in[n], \pair{i,j}<\pair{i',j'}, ((i,j)\notin E \vee  ({i'},{j'})\notin E)\}$, and $\{(\pair{i,j},\pair{1,j'}) \mid i,j,j'\in[2,n]_{\integer}\}$.
Let $b=c+w_{*}+1$.

We define the desired instance $y$ as $y= ((1^b,1^{a_{k_1}},1^{a_{k_2}},\ldots,1^{a_{k_{m}}}),EXC)$. For this instance $y$, we can prove that $x\in \mathrm{TO\mbox{-}EWPP}$ iff  $y\in \mathrm{UKEXC}$.
\end{proofsketch}

\section{Lattice Problems}\label{sec:lattice-theory}

Let us discuss lattice problems. A \emph{lattice} is a set of integer linear combinations of basis vectors. Here, we deal only with the case where bases are taken from $\integer^n$ for a certain $n\in\nat^{+}$.
The decision version of the \emph{closest vector problem} (CVP) is known to be $\np$-complete \cite{Emd81}.
We then consider a variant of $\mathrm{CVP}$.
To fit into our setting of unary-form integers,
a simple norm notion of the \emph{max norm} $\|\cdot\|_{\infty}$ from  Section \ref{sec:numbers}.
In a syntactic similarity, we further introduce the notation $\|v\|_{\min}$ to express $\min\{|v_i|: i\in[n]\}$ for
any real-valued vector $v=(v_1,v_2,\ldots,v_n)$. Notice that $\|v\|_{\min}$ does not serve as a true ``distance''.
For convenience, $\LL(v_1,v_2,\ldots,v_m)$ denotes the lattice spanned by a given set  $\{v_1,v_2,\ldots,v_m\}$ of basis vectors.

\ms
{\sc Unary Max-Norm Closest Vector Problem}  ({\sc UCVP}$_{\max}$):
\renewcommand{\labelitemi}{$\circ$}
\begin{itemize}\vs{-1}
  \setlength{\topsep}{-2mm}%
  \setlength{\itemsep}{1mm}%
  \setlength{\parskip}{0cm}%

\item {\sc Instance:} $1^b$ for a positive integer $b$, a tuple $(1^{\pair{v_i[1]}}, 1^{\pair{v_i[2]}}, \ldots, 1^{\pair{v_i[n]}})$ for a set $\{v_1,v_2,\ldots,v_m\}$ of lattice bases with $v_i = (v_i[1],v_i[2],\ldots,v_i[n])\in \integer^n$, and a tuple $(1^{\pair{x_0[1]}},1^{\pair{x_0[2]}}, \ldots, 1^{\pair{x_0[n]}})$ for a target vector $x_0=(x_0[1],x_0[2],\ldots,x_0[n])\in\integer^n$.

\item {\sc Question:} is there a lattice vector $w$ in $\LL(v_1,v_2,\ldots,v_m)$ such that the max norm  $\|w-x_0\|_{\infty}$ is at most $b$?
\end{itemize}

In the above definition of $\mathrm{UCVP}_{\max}$, we replace $\|\cdot\|_{\infty}$ by $\|\cdot\|_{min}$ and then obtain
$\mathrm{UCVP}_{\min}$.

For simplicity, in what follows, we intend to write $\bar{v}$ for $(1^{\pair{v[1]}}, 1^{\pair{v[2]}}, \ldots, 1^{\pair{v[n]}})$ if $v=(v[1],v[2],\ldots,v[n])$.


\begin{proposition}\label{SU2PART-to-UCVP}
$\mathrm{SUK}\leq^{\mathrm{L}}_{m} \mathrm{UCVP}_{\max}$.
\end{proposition}

\begin{proofsketch}
In a similar way to obtaining $\mathrm{SUK}$ from $\mathrm{UK}$, we can introduce a variant of $\mathrm{U2PART}$, called the \emph{simultaneous unary 2 partition problem} (SU2PART). It is possible to prove that  $\mathrm{SUK}\leq^{\dl}_{m}\mathrm{SU2PART}$. It therefore suffices to verify that $\mathrm{SU2PART}\leq^{\dl}_{m}\mathrm{UCVP}_{\max}$.

Let $x= (a_1,a_2,\ldots,a_m)$ with $a_j=(1^{a_{j1}},1^{a_{j2}},\ldots,1^{a_{jn}})$ for any $j\in[m]$
be any instance of $\mathrm{SU2PART}$. Let $d_j =\sum_{i\in[n]}a_{ji}$ for each $j\in[m]$ and set $d_{max}=\max_{j\in[m]}\{d_j\}$. Given any $j\in[m]$ and $i\in[n]$, we further set $a'_{ji}= d_{max}a_{ji}$ and $d'_j=d_{max}d_j$.
Let us define $n$ vectors $v_1,v_2,\ldots,v_n$ as $v_1=(a'_{11},a'_{21}, \ldots, a'_{m1},2,0,0, \ldots,0)$, $v_2=(a'_{21},a'_{22},\ldots, a'_{2n}, 0,2,0,\ldots,0)$, $\ldots$, $v_n=(a'_{n1},a'_{n2},\ldots, a'_{nm}, 0,0,0,\ldots,0,2)$.
Moreover, we set $x_0=(\floors{d'_1/2}, \floors{d'_2/2}, \ldots, \floors{d'_m/2}, 1,1,\ldots,1)$ and $b=1$.
We then define $y$ to be $(1^{b},\bar{v}_1,\bar{v}_2,\ldots,\bar{v}_n,\bar{x}_{0})$.
Clearly, $y$ is an instance of $\mathrm{UCSP}_{\max}$.
It then follows that $x\in \mathrm{U2PART}$ iff $y\in \mathrm{UCVC}_{\max}$.
\end{proofsketch}


\begin{lemma}\label{UCVP-NL}
(1) $\mathrm{UCVP}_{\max} \in \mathrm{1t2}\Sigma_2\mathrm{PDCA}$.
(2) $\mathrm{UCVP}_{\min} \in \mathrm{1t2NCA}$.
\end{lemma}

It is not clear that $\mathrm{UCVP}_{\max}$ belongs to $\mathrm{1t2NPDCA}$ or even $\p$.


Next, we look into another relevant problem, known as the \emph{shortest vector problem}. We consider its variant.

\ms
{\sc Unary Max-Norm Shortest Vector Problem}  ({\sc USVP}$_{\max}$):
\renewcommand{\labelitemi}{$\circ$}
\begin{itemize}\vs{-1}
  \setlength{\topsep}{-2mm}%
  \setlength{\itemsep}{1mm}%
  \setlength{\parskip}{0cm}%

\item {\sc Instance:} $1^b$ with $b\in\nat^{+}$
    and a tuple $(1^{\pair{v_i[1]}}, 1^{\pair{v_i[2]}}, \ldots, 1^{\pair{v_i[n]}})$ for
    a set  $\{v_1,v_2,\ldots,v_m\}$ of lattice bases with  $v_i = (v_i[1],v_i[2],\ldots,v_i[n])\in \integer^n$.

\item {\sc Question:} is there a ``non-zero'' lattice vector $w$ in $\LL(v_1,v_2,\ldots,v_m)$ such that the max norm  $\|w\|_{\infty}$ is at most $b$?
\end{itemize}

Similarly, we can define $\mathrm{USVP}_{\min}$ by replacing $\|\cdot\|_{\infty}$ with $\|\cdot\|_{\min}$.
In a way similar to prove Lemma \ref{UCVP-NL}, we can prove that $\mathrm{USVP}_{\max}\in \mathrm{1t2}\Sigma_2\mathrm{PDCA}$. Moreover, the following relations hold.

\begin{lemma}\label{USVP-tt-UCVP}
$\mathrm{USVP}_{\min} \Lttreduces \mathrm{UCVP}_{\min}$ and $\mathrm{USVP}_{\max} \Lttreduces \mathrm{UCVP}_{\max}$.
\end{lemma}

We do not know if $\mathrm{USVP}_{\min}\equiv^{\dl}_{tt} \mathrm{UCVP}_{\min}$ and $\mathrm{USVP}_{\max}\equiv^{\dl}_{tt} \mathrm{UCVP}_{\max}$.

\let\oldbibliography\thebibliography
\renewcommand{\thebibliography}[1]{%
  \oldbibliography{#1}%
  \setlength{\itemsep}{-1pt}%
}
\bibliographystyle{alpha}

\begin{thebibliography}{99}
{\small

\bibitem{CH88}
Cho, S., Huynh, D.T.: On a complexity hierarchy between L and NL. Inform. Process. Lett. 29, 177--182 (1988)

\bibitem{CHS74}
Constable, R.L., Hunt III, H.B., Sahni, S.: On the computational complexity of scheme equivalence. Report No. 74-201, Dept. of Computer Science, Cornell University (1974)

\bibitem{Coo79}
Cook, S.A.: Deterministic CFLs are accepted simultaneously in polynomial time and log squared space. In the Proceedings of the 1lth Annual ACM Symposium on Theory of Computing, pp. 338--345 (1979)

\bibitem{Coo85}
Cook, S.A.: A taxonomy of problems with fast parallel algorithms. Inform. Control 64 (1985) 2--22.

\bibitem{GJ79}
Garey, M.R.,  Johnson, D.S.: Computers and Interactability: A Guide to the Theory of NP-Completeness, W. H. Freeman and Company, New York (1979)

\bibitem{GS66}
Ginsburg, G., E. H. Spanier, E.H.: Finite-turn pushdown automata. SIAM J. Control 4, 429--453 (1966)

\bibitem{Jen95}
Jenner, B.: Knapsack problems for NL. Inform. Process. Lett. 54,
169--174 (1995)

\bibitem{JK89}
Jenner, B.,  Kirsg, B.: Characterizing the polynomial hierarchy by alternating pushdown automata.
RAIRO--Theoret. Inform. App. 23, 87--99 (1989)

\bibitem{JLL76}
Jones, N.D., Lien, Y.E., Laaser, W.T.:  New problems complete for
nondeterministic log space. {Math. Systems Theory} 10, 1--17 (1976)

\bibitem{Kar72}
Karp, R.M.: Reducibility among combinatorial problems. In: R. E. Miller and J. W. Thatcher (eds.), Complexity of Computer Computations. Plenum Press, New York, pp. 85--103 (1972)

\bibitem{Min67}
Minsky, M.L.: Computation: Finite and Infinite Machines.
Prentice-Hall, Englewood Cliffs, NJ (1967)

\bibitem{MS80}
Monien, B.,  Sudborough, I.H.: Foraml language theory. In: Formal Language Theory, R. V. Book (ed), Academic Press, New York (1980).

\bibitem{Rein08}
Reingold, O.: Undirected connectivity in log-space. J. ACM 55,
article 17 (2008)

\bibitem{SM73}
Stockmeyer, L.J.,  Meyer, A.R.: Word problems requiring exponential time. In the Proc. of the 5th Ann. ACM Symp. on Theory of Computing, pp. 1--9 (1973)

\bibitem{Sud78}
Sudborough, I.H.: On the tape complexity of determinsitic context-free languages. J. ACM 25, 405--414 (1978)

\bibitem{Emd81}
van Emde Boas, P.: Another NP-complete partition problem and the complexity of computing short vectors in a lattice. Report No. 81-04, Department of Mathematice, University of Amsterdam (1981)

\bibitem{Yam23a}
T. Yamakami. Unambiguous and co-nondeterministic computations of finite automata and pushdown automata families and multiple counters. Manuscript,  2023.

}
\end{thebibliography}

\end{document}